\documentclass{elsart}

\usepackage{graphicx}

\begin{document}

\newcommand{\snn}{\sqrt{s_{NN}}}
\newcommand{\seff}{\sqrt{s_{\rm eff}}}
\newcommand{\s}{\sqrt{s}}
\newcommand{\pp}{pp}
\newcommand{\pbarp}{\overline{p}p}
\newcommand{\qbarq}{\overline{q}q}
\newcommand{\epem}{e^+e^-}

\newcommand{\nhit}{N_{hit}}
\newcommand{\npp}{n_{pp}}
\newcommand{\nch}{N_{ch}}
\newcommand{\np}{N_{part}}
\newcommand{\ns}{N_{spec}}
\newcommand{\ntot}{\langle\nch\rangle}
\newcommand{\avenp}{\langle\np\rangle}
\newcommand{\npB}{N_{part}^B}
\newcommand{\nc}{N_{coll}}
\newcommand{\avenc}{\langle\nc\rangle}
\newcommand{\halfnp}{\langle\np/2\rangle}
\newcommand{\etap}{\eta^{\prime}}
\newcommand{\as}{\alpha_{s}(s)}
\newcommand{\etazero}{\eta = 0}
\newcommand{\etaone}{|\eta| < 1}
\newcommand{\dndeta}{d\nch/d\eta}
\newcommand{\dndetazero}{\dndeta|_{\etazero}}
\newcommand{\dndetaone}{\dndeta|_{\etaone}}
\newcommand{\dndetanp}{\dndeta / \halfnp}
\newcommand{\dndetaonp}{\dndeta / \np}
\newcommand{\dndetazeronp}{\dndetazero / \halfnp}
\newcommand{\dndetaonenp}{\dndetaone / \halfnp}
\newcommand{\ratio}{\ntot/\halfnp}
\newcommand{\nee}{N_{\epem}}
\newcommand{\nhh}{N_{hh}}
\newcommand{\nubar}{\overline{\nu}}

\begin{frontmatter}

\title{Inclusive Pseudorapidity Distributions in p(d)+A Collisions Modeled With Shifted Rapidity Distributions}

\author{P. Steinberg}
\address{Chemistry Department\\
Brookhaven National Laboratory, Upton, NY 11973, USA}
\date{\today}

\begin{abstract}
We study the recent PHOBOS data on the pseudorapidity density
of inclusive charged particles in centrality-binned d+Au
collisions at $\snn=200$ GeV.
It appears that one can understand the increasing forward-backward 
asymmetry in the data by assuming that the entire
distribution ``shifts'' backwards in rapidity according 
to the initial-state kinematics, while the total multiplicity scales 
linearly with $\np/2$.
Two models are explored, both of which achieve a reasonable
description of the available data for rapidities sufficiently
far from the projectiles ($|\eta|<3$ at the top  RHIC energy).
One model uses PYTHIA as the underlying
distribution, to allow a straightforward mapping of a  
rapidity shift back to pseudorapidity space.  
The other model is
a simple analytic calculation based on functions inspired by
Landau's hydrodynamical model.
The apparent success of these simple pictures to describe
the bulk of particle production over $|\eta|<3$ suggests that
collective effects may be present even in the ``small'' systems
created in p+p and d+Au reactions, active over the full
rapidity range.  
The relationship between this and other theoretical approaches
are discussed.
\end{abstract}
\end{frontmatter}
Recent PHOBOS data \cite{Back:2003hx} have shown the pseudorapidity
distribution of charged particles for minimum-bias d+Au collisions
at the full RHIC energy $\snn =$ 200 GeV as well as the centrality
dependence of the distributions.
In this work, we will show that the bulk particle production
in p+A and d+Au collisions over a range of energies can be
described by a very simple model with only two assumptions:
\begin{itemize}
\item The total multiplicity in p+A and d+Au collisions is 
linear with the number of participant pairs ($\np/2$) relative
to p+p collisions (similar to the 
``wounded-nucleon model''~\cite{Bialas:1976ed,Elias:1978ft}.
\item The angular distribution in p+A and d+Au collisions are
approximately the same as p+p collisions (if one avoids the
very far-forward rapidity regions) but in a ``shifted''
reference frame given by the center-of-mass system defined by
the participants in the smaller and larger projectiles.
\end{itemize}

\begin{figure}[t]
\begin{center}
\includegraphics[width=8cm]{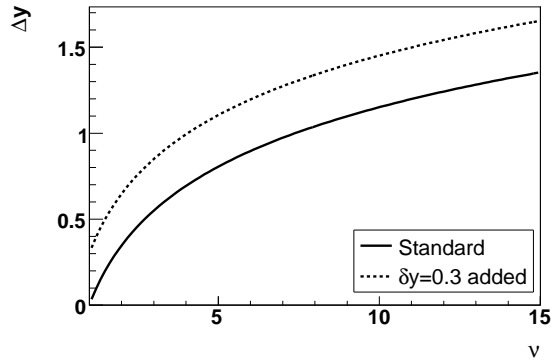}
\end{center}
\caption{
The center-of-mass rapidity shift $\Delta y$ vs. $\nu$,
the ratio of number of participants in the nucleus to that
in the smaller projectile.  The same function shifted up
by 0.3 units is also shown,to account for ``extra'' rapidity
shifts due to the geometry or to transverse dynamics.
}
\label{figure1}
\end{figure}

Simple kinematics gives the result that the rapidity of the
center-of-mass (CM) frame in a p+A collision, where the proton
interacts with a ``tube'' of $\nu$ nucleons in the nucleus, is
\[
\label{ypa}
\Delta y_{p+A} = \frac{1}{2}\ln\left(\nu\right)
\]
In a d+Au collision, one can approximate the geometry to be
the independent collision of the proton and neutron with 
half of the participants in the gold nucleus, or a full
collision of the 2 nucleons in the deuteron with all of the
participants in the nucleus.
In either case,
\[
\label{yda}
\Delta y_{d+Au} \sim \frac{1}{2}\ln\left(\frac{N_{Au}}{N_{d}}\right)
\]
In general, these two definitions lead to equivalent physics
conclusions, with $\nu\equiv N_{Au}/ N_{d}$.

According to the above assumptions, the produced particles
should emerge with a rapidity distribution similar to the
underlying p+p distribution, but shifted by $\Delta y$
relative to the original p+p CM frame.  However, it should
be noted that if the relevant physics involves rescaling
$x_F=2p_T/\sqrt{s}$, where $dN/dx_F$ is universal
with energy for a given initial state geometry, then there might
be an additional modification to the rapidity distribution,
due to the transverse momentum of the emitted particles.  This
is easily seen by the approximate expression relating $y$ and $x_F$
(good for $y$ away from 0):
\[
y = y_{b} + \ln(x_F) - \frac{1}{2}\ln\left(\frac{m_T^2}{M_P^2}\right)
\]
Without a full knowledge of the longitudinal and transverse distributions,
it is difficult to estimate the complete effect of the dynamical evolution
on the final state rapidity distributions.  
One might also consider some additional effects from reinteractions
with the spectator matter from the nucleus, supplying the system with
some extra longitudinal momentum in the backwards (A) hemisphere. 
Despite these uncertainties, 
it will be assumed that any additional modifications to the longitudindal
shift (e.g. from effects associated with transverse dynamics or
spectator interactions), can be incorporated into a constant
extra contribution to $\Delta y$, hereby labeled $\delta y$.
The basic relationship between the initial-state geometry and
the rapidity shift is shown in Fig.\ref{figure1}, with and
without an extra shift of $\delta y=0.3$.

\begin{figure}[t]
\begin{center}
\includegraphics[width=8cm]{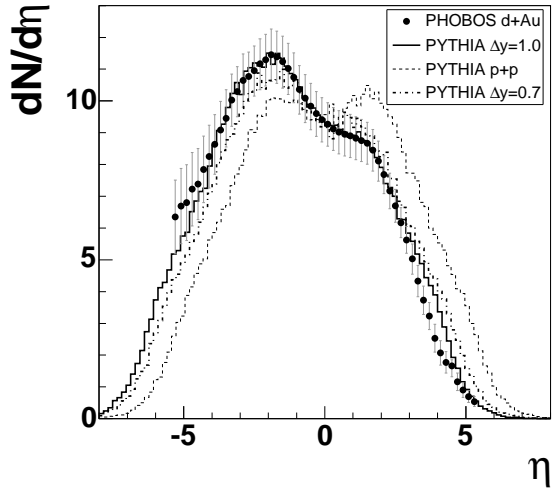}
\end{center}
\caption{
PHOBOS $\dndeta$ distribution for d+Au collisions compared with
PYTHIA $\pp$ simulations scaled up by $\np/2$ and with the
rapidity of each particle shifted by $\Delta y = $0.7 and 1.
}
\label{figure2}
\end{figure}

The simplest application of the approach described here can be
performed with the PHOBOS minimum-bias d+Au data, with 
$\nu$ estimated to be approximately 4.  This data, shown 
in Fig.~\ref{figure2}, displays a clear rapidity asymmetry,
with a large peak seen near $\eta=-2$ and an intriguing 
``plateau'' near $\eta=1$.  These features are often interpreted
in terms of the independent fragmentation of the deuteron
and nucleus in the forward and backward rapidity regions
respectively.  However, we can use the procedure outlined above to
construct d+Au distributions from p+p, using 
PYTHIA 6.161~\cite{Sjostrand:2000wi}.
First one scales
up the PYTHIA p+p $\dndeta$ distribution up by a factor of $\np/2
\sim 4$, indicated by a dotted line.  This gives a
distribution with similar integral to the PHOBOS data, but
one which is clearly higher in the forward region and
lower in the backward region.  Then we apply a rapidity
shift $\Delta y$ to each particle in the PYTHIA simulation
and then recalculate $\dndeta$ for $y\rightarrow y-\Delta y$.
Using the straightforward formula, Eq.~\ref{yda}, we would
get $\Delta y=0.7$, shown by the dot-dashed line.  Already
we see that this result describes the basic features of the
PHOBOS data, including the peak at $\eta=-2$, which is now
seen clearly as a kinematic effect resulting from using 
pseudorapidity ($\eta$) instead of rapidity ($y$).  However, it is also
shown that assuming $\delta y=0.3$, giving a total $\Delta y=1.0$,
gives a somewhat better fit with the data over the measured range.

This exercise shows that the simple model proposed above can in
fact achieve a satisfactory quantitative agreement with 
the data. However, a slight modification ($\delta y=0.3$)
can improve this agreement substantially.  The next
question then how far one can test these ideas by varying
the collision centrality, which changes $\nu$.  
Fortunately, the PHOBOS collaboration has
also published data comparing d+Au data binned in centrality
bins (5 bins, each containing 20\% of the total inelastic
d+Au cross section)\cite{Back:2004mr}, and presented as a ratio compared to
p+p data from UA5, i.e. the variable
\[
\label{rdapp}
R_{\eta}=\frac{dN_{d+Au}/d\eta}{dN_{p+p}/d\eta}.
\]
The BRAHMS collaboration has presented d+Au data binned in centrality
bins~\cite{Arsene:2004cn}, 
but their centrality method (using particles emitted into
$|\eta|<2.2$) creates severe biases on the final distributions.
This makes direct comparisons with the models shown here quite difficult
without substantial simulation work to remove them.

\begin{figure*}[t]
\begin{center}
\includegraphics[width=6.5cm]{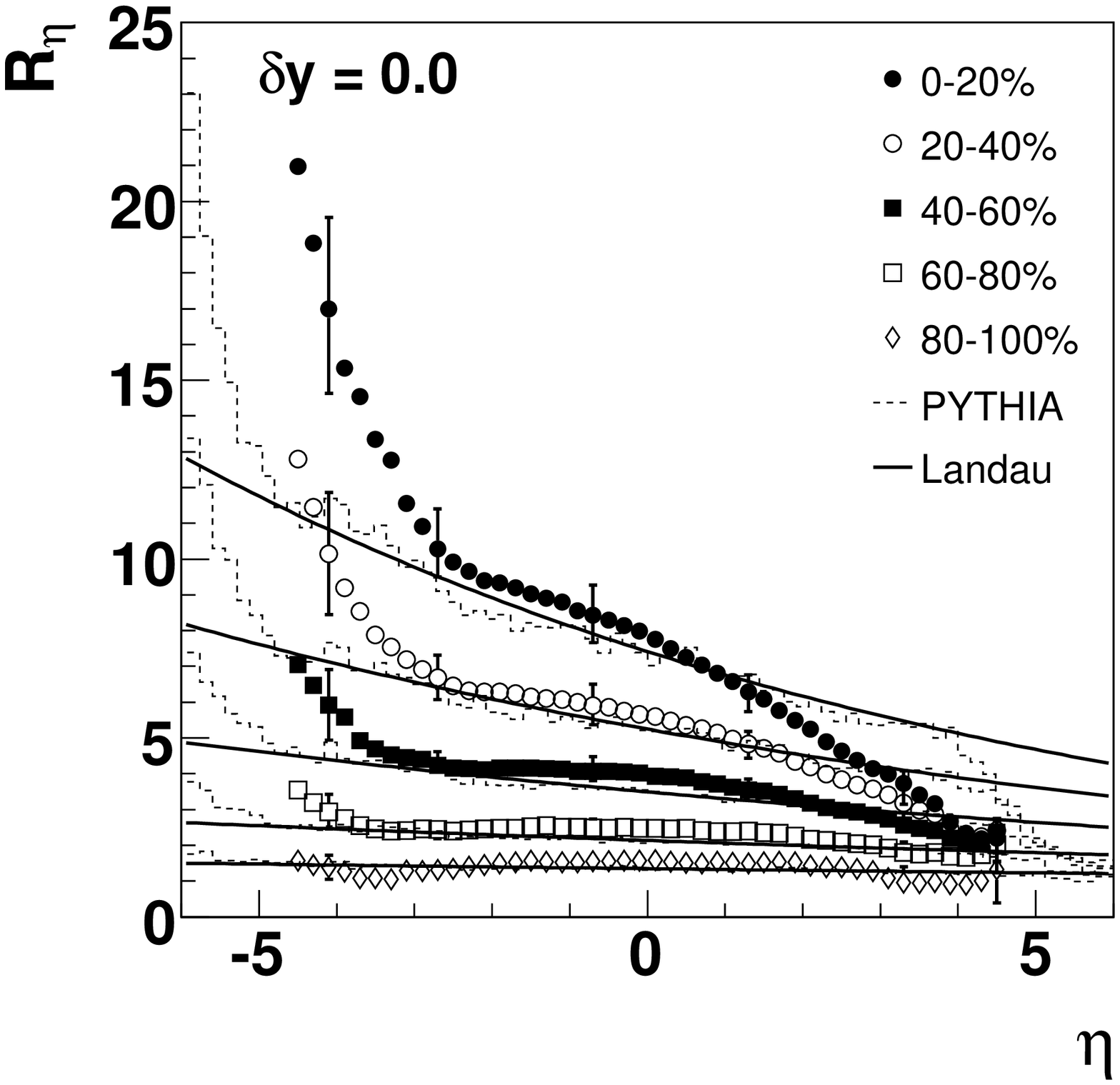}
\includegraphics[width=6.5cm]{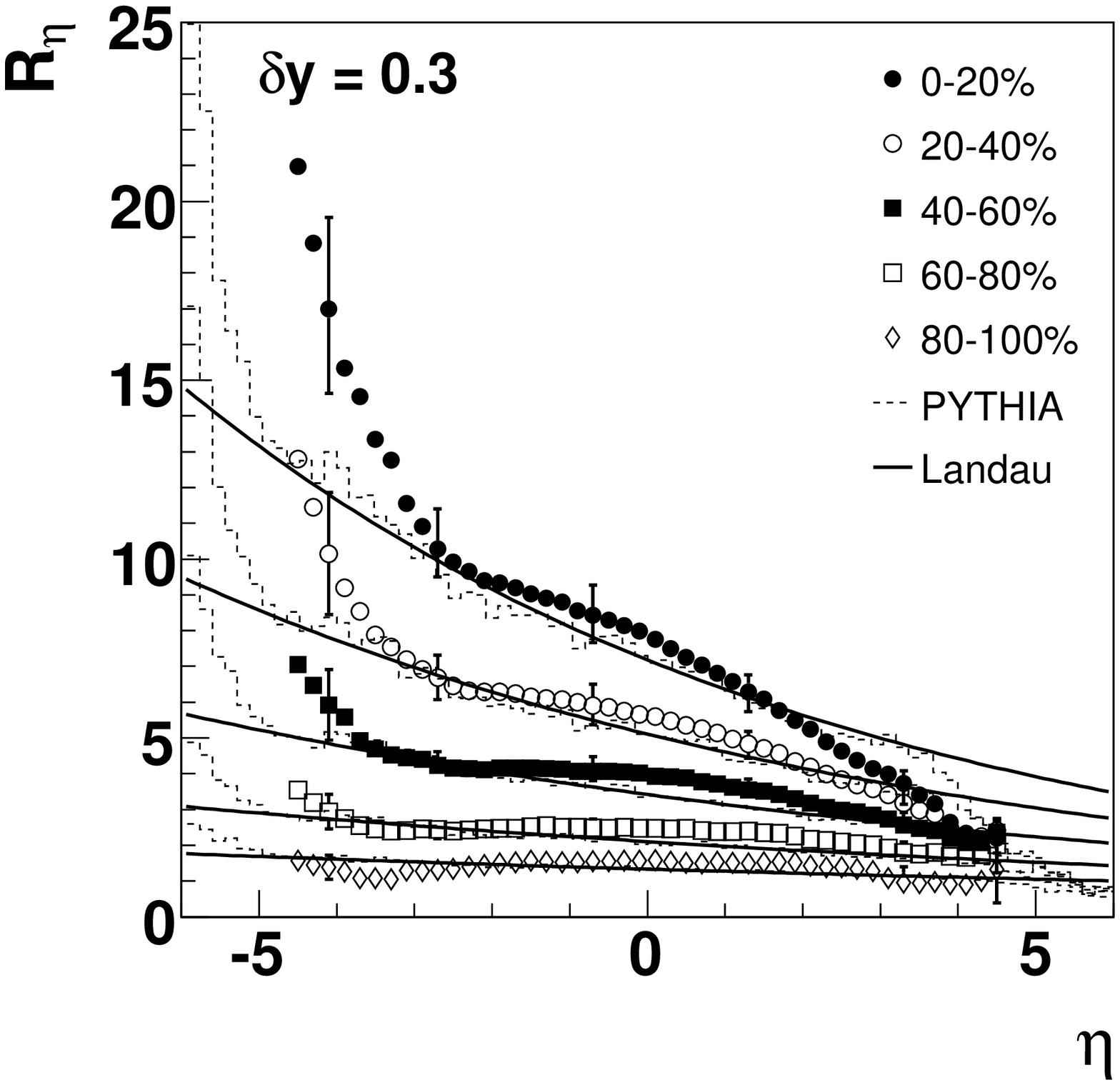}
\end{center}
\caption{
The ratio of d+Au and p+p pseudorapidity distributions is shown 
for a range of centralities in d+Au (corresponding to Ref. 
\cite{Back:2004mr}).  
The data is compared to two calculations, one using
PYTHIA (dotted histograms) and the using a simple analytic model inspired by
Landau's hydrodynamical model (solid line).  The left and
right panels are for $\delta y=0.0$ and $\delta y=0.3$, respectively.
}
\label{figure3}
\end{figure*}

We have made theoretical calculations
of $R_\eta$ to compare with the PHOBOS data
in two different ways.  One is to use the prescription above
($\np/2$ scaling and a $\Delta y$ shift)
on PYTHIA calculations, which has the advantage of allowing
a full calculation involving the translation from $y$ to
$\eta$, to see the effect of the Jacobian.  In Fig.~\ref{figure3},
we show these PYTHIA calculations as solid histograms, which
describe the data reasonably well within $|\eta|<3$.
The two panels both illustrate that the Jacobian essentially
cancels in the ratio of d+Au and p+p.  However, in the comparison of
$\delta y=0$ (which is a parameter-free calculation) and $\delta y=0.3$, 
it is observed that the non-zero $\delta y$ gives a somewhat better
description of the data.  These results thus support and
extend the one shown in Fig.~\ref{figure2}.

The other model is
an analytic calculation based on Landau's hydrodynamics,
as presented by Carruthers and others~\cite{Carruthers:dw}.  
In this approach,
the rapidity distribution in p+p is a Gaussian with width
$\sigma^2 = (1/2)\ln(s/4m^2)$, a function particularly simple
to scale by $\np/2$ and shift by $\Delta y$.  One can then
calculate $R_y$, which is the same as $R_\eta$ but in rapidity
space:
\begin{eqnarray}
R_y & = & \frac{\np}{2} \frac{e^{-(y+\Delta y)^2/2\sigma^2}}{e^{-y^2/2\sigma^2}} \\
\nonumber & = & \frac{\np e^{-\Delta y^2/2\sigma^2}}{2} e^{-(y \Delta y)/\sigma^2}
\end{eqnarray}
which gives a exponential dependence of $R_{y}$ with $y$.  For simplicity,
we will approximate $y$ in the final expression by $\eta$, which
appears justified by the PYTHIA results, as the $\eta$-$y$ transformation
does not seem to have a great effect on $R_{\eta}$.  These results are
shown in Fig.~\ref{figure3} as solid lines, and also give an adequate
representation of the data, within $|\eta|<3$.

\begin{figure*}[t]
\begin{center}
\includegraphics[width=14cm]{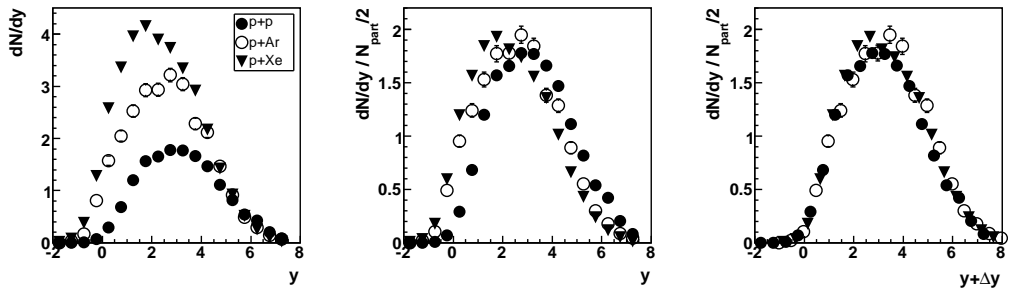}
\end{center}
\caption{
left) NA5 data on dN/dy for inclusive produced charged particles, 
shown for three different targets.
center) The same data scaled by the number of participant pairs.
right) The same data as a function of $y+\Delta y$, with $\delta y=0.3$
(except for p+p).
}
\label{figure4}
\end{figure*}

To test the applicability of the basic approach outlined here to
lower-energy reactions, the procedure outlined above has been applied to
the NA5 data set of $dN/dy$ for inclusive produced charged particles
measured in 200 GeV protons incident
on hydrogen, argon and xenon targets~\cite{DeMarzo:1982rh}.  
Varying the target nucleus
varies the number of collisions and allows us to test the
predicted formula for the rapidity shift, as well as $\np$ scaling.
The first panel of Fig.~\ref{figure4} shows the raw data on dN/dy,
the {\it rapidity} distribution of inclusive charged particles assuming
the pion mass.  The second panel shows these distributions divided by
$\np/2$, where $\np = \nu +1$ and the values of $\nu$ are taken from
Ref. \cite{DeMarzo:1982rh}.  The third panel shows the scaled distributions 
shifted to the right using the formula derived above (with $\delta y= 0.3$).  
After these transformations,  a reasonable agreement of the distributions
is observed over a large rapidity range.  
It is also observed that the p+A collisions
are quite symmetrical over the full range relative to the centroid, 
even when shifted by up to a unit in rapidity.  

Rapidity shifts in the context of ``fireball'' desriptions of
heavy ion collisions have
been discussed since the late 1970's\cite{Gosset:1988na}.
However, as far as the author is aware,
the shifted rapidity distribution for p+A collisions
has only been incorporated explicitly into one concrete
physical model,
the Collective Tube Model (CTM) \cite{Afek:1976in}.
The CTM modelled p+A collisions as being the collision of a
proton with another particle of the same size as a proton, but
with a mass and momentum $\nu$ times larger.
This also predicted a shift of all
produced particles by $\ln(A^{1/6})$ (compatible with the simple
calculation shown here given that $\nu \propto A^{1/3}$).
However, to rigorously
conserve longitudinal momentum as well as total energy, this model
also predicted substantially fewer particles than observed in a large
number of experiments, and a centrality dependence $(\propto \np^{1/4})$
which disagrees with the $\np$-scaling observed in all existing p+A data.  

Shifted rapidity distributions have also been discussed in the context
of Landau's hydrodynamical model~\cite{Landau:1953gs}.
In Landau's approach, the incoming projectiles thermalize in a volume
reduced by the Lorentz contraction, which increases with energy as fast
as the CM energy itself.  The angular distributions are then the
consequence of a rapid longitudinal expansion, which can be calculated
in the context of relativistic hydrodynamics.
Landau and Belenkij~\cite{Belenkij:1956cd}
performed a detailed calculation on particle 
production in p+A collisions, and found that under reasonable
conditions ($\nu<4$), that the total multiplicity should scale
with $\np$ and that the angular distributions should differ
``only slightly'' from that found in nucleon-nucleon collisions.
Carruthers~\cite{Carruthers:dw}
suggested that asymmetric collisions should lead to ``shifted
gaussians'', but detailed phenomenological studies were not made.
The salient point to distinguish the Landau approach from the CTM
approach is that in the former, one would not assume all of the
energy in the tube of $\nu$ nucleons to be compressed into a volume 
the size of a (Lorentz-contracted) nucleon.  Rather, it has a
longitudinal extent approximately $\nu$ times a nucleon, such that
it's thermalized energy density is not dissimilar from a p+p 
collision.  

The most important issue is that one should not expect the full
inclusive distribution to be shifted by $\Delta y$, as this would
manifestly violate energy-momentum conservation if extended all
the way to beam rapidities, not in the least when one shifts the
leading particles on the nucleus side to a larger rapidity than
they had before the interaction.  Thus, it is perhaps not surprising that
the two models presented here fail at large rapidities on both
the projectile and target sides, as seen in Fig.~\ref{figure3}. 
Of course, the details of these deviations may well
give useful information either about the thermalization
process, or even about the particular issues involved with 
hydrodynamic evolution near to kinematic boundaries.

Finally, we compare the model sketched here with several existing
in the literature.  In effect, we present an alternative explanation to
what had been suggested by Brodsky {\it et al.}\cite{Brodsky:1977de} 
to be the result
of the multiple-collision process and overlapping ``strings''
between the colliding nucleons.  They predicted a linear decrease
of $R_{\eta}$ with $\eta$, while the models here predict an approximately
exponential decrease.  There is also a substantive difference between
this approach and that advocated recently by Bialas and Czyz, motivated
by a successful fit of the same data with an updated ``wounded nucleon'' 
model~\cite{Bialas:2004su}.  In particular, the physical picture advocated
here suggests that it is the collective system which expands in a shifted frame
to make the final state rapidity distribution, rather than the independent
fragmentation of the incoming nucleons.  This avoids the situation,
detailed by Bialas and Czyz, 
that the forward-going nucleons can fragment into fast-moving particles
heading in the opposite direction.

Of course, the limited precision of both the
data and the theory precludes any decisive choice of any of
these models at this moment.  
The physical picture outlined in this work has some appeal 
having essentially no free parameters except
the residual shift ($\delta y$), and requires no additional factorization
of p+p collisions into two fragmentation regions, as suggested by 
``wounded nucleon'' approaches.
However, it should be noted
that explaining this ``shift'' in the overall rapidity distribution
by a collective response of the bulk of the produced particles to the shift
in the center of mass in the initial state is not immediately consistent with a
parton model approach.  To achieve similar results, one must introduce
``shadowing'' and ``anti-shadowing'' of the bound-nucleon structure 
functions to achieve similar effects in the rapidity distributions.
How this should preserve ``wounded nucleon'' scaling in a natural way
is not clear from the perturbative approach,
perhaps leaving room for simpler, heuristic models to point to issues
where additional theoretical guidance will be needed in the future.

As a final note: as this manuscript was being completed, the author 
became aware of similar approaches to understanding the RHIC d+Au
data, based on ``shifting'' the bulk distributions in 
rapidity\cite{Wolschin:2005by} 
or pseudorapidity\cite{Jeon:2003nx,Putschke:2005zz}.  
It should be noted that there are basic differences.  For example,
Jeon et al \cite{Jeon:2003nx} compare d+Au data with
Au+Au data, applying scaling and shifts in $dN/d\eta$ and shifts in 
pseudorapidity space, where the work here considers shifts in
rapidity and no rescaling of the multiplicity beyond participant
scaling between d+Au and p+p data.
Still, it is intriguing to see similar ideas to those presented here
appearing independently in both theoretical and experimental works.

This work was supported under U.S. DOE grant DE-AC02-98CH10886.
The author acknowledges insightful discussions with Mark Baker, Wit Busza
and Rachid Nouicer.


\begin{thebibliography}{50}
\bibitem{Back:2003hx} B.~B.~Back {\it et al.},  Phys.\ Rev.\ Lett.\  {\bf 93}, 082301 (2004).
\bibitem{Bialas:1976ed} A.~Bia\l as, B.~Bleszy\'{n}ski and W.~Czy\.{z}, Nucl.\ Phys.\ {\bf B111}, 461 (1976).
\bibitem{Elias:1978ft} J.~E.~Elias {\it et al.},  Phys.\ Rev.\ Lett.\  {\bf 41}, 285 (1978).
\bibitem{Sjostrand:2000wi}  T.~Sjostrand {\it et al.}, Comput.\ Phys.\ Commun.\  {\bf 135}, 238 (2001).
\bibitem{Back:2004mr} B.~B.~Back {\it et al.}, arXiv:nucl-ex/0409021.
\bibitem{Arsene:2004cn} I.~Arsene {\it et al.}, Phys.\ Rev.\ Lett.\  {\bf 94}, 032301 (2005).
\bibitem{Carruthers:dw} P.~Carruthers, Annals N.Y.Acad.Sci. 229, 91 (1974).
\bibitem{DeMarzo:1982rh} C.~De Marzo {\it et al.},  Phys.\ Rev.\ D {\bf 26}, 1019 (1982).
\bibitem{Gosset:1988na}  J.~Gosset, J.~I.~Kapusta and G.~D.~Westfall, Phys.\ Rev.\ C {\bf 18}, 844 (1978).
\bibitem{Afek:1976in} Y.~Afek, G.~Berlad, A.~Dar and G.~Eilam, TECHNION-PH-76-87 (1976).
\bibitem{Landau:1953gs} L.~D.~Landau, Izv.\ Akad.\ Nauk Ser.\ Fiz.\  {\bf 17}, 51 (1953).
\bibitem{Belenkij:1956cd} S.~Z.~Belenkij and L.~D.~Landau, Nuovo Cim.\ Suppl.\  {\bf 3S10}, 15 (1956)  [Usp.\ Fiz.\ Nauk {\bf 56}, 309 (1955)].
\bibitem{Brodsky:1977de}  S.~J.~Brodsky, J.~F.~Gunion and J.~H.~Kuhn,  Phys.\ Rev.\ Lett.\  {\bf 39}, 1120 (1977).
\bibitem{Bialas:2004su} A.~Bialas and W.~Czyz,  Acta Phys.\ Polon.\ B {\bf 36}, 905 (2005).
\bibitem{Wolschin:2005by} G.~Wolschin, M.~Biyajima, T.~Mizoguchi and N.~Suzuki, arXiv:hep-ph/0503212.
\bibitem{Jeon:2003nx} S.~Jeon, V.~Topor Pop and M.~Bleicher, Phys.\ Rev.\ C {\bf 69}, 044904 (2004)
\bibitem{Putschke:2005zz} J.~Putschke,  J.\ Phys.\ Conf.\ Ser.\  {\bf 5}, 37 (2005).  J.~Putcshke, PhD Thesis.
\end{thebibliography}
\end{document}